\documentstyle[12pt,titlepage]{article}
\input epsf.tex

\renewcommand{\theequation}{\thesection.\arabic{equation}}
\font\small=cmr8 scaled \magstep0
\font\grande=cmr10 scaled \magstep4
\font\medio=cmr10 scaled \magstep2
\outer\def\beginsection#1\par{\medbreak\bigskip
      \message{#1}\leftline{\bf#1}\nobreak\medskip\vskip-\parskip
      \noindent}

\def\laq{\raise 0.4ex\hbox{$<$}\kern -0.8em\lower 0.62
ex\hbox{$\sim$}}
\def\gaq{\raise 0.4ex\hbox{$>$}\kern -0.7em\lower 0.62
ex\hbox{$\sim$}}

\def \la {\lambda}

\def \b {\beta}

\def \om {\omega}
\def \Om {\Omega}
\def \noi {\noindent}

\def\sqr#1#2{{\vcenter{\hrule height.#2pt\hbox{\vrule width.#2pt
height#1pt \kern#1pt\vrule width.#2pt}\hrule height.#2pt}}}

\def\lsim{\mathrel{\rlap{\lower4pt\hbox{\hskip1pt$\sim$}}
    \raise1pt\hbox{$<$}}}         
\def\gsim{\mathrel{\rlap{\lower4pt\hbox{\hskip1pt$\sim$}}
    \raise1pt\hbox{$>$}}}         

\def\b{\beta}
\def\d{\delta}

\def\n{\eta}

\def\w{\omega}

\def\bq{\begin{equation}}
\def\eq{\end{equation}}
\def\brr{\begin{eqnarray}}
\def\err{\end{eqnarray}}
\def\ba{\left(\begin{array}}
\def\ea{\end{array}\right)}

\def\ba{\left(\begin{array}}
\def\ea{\end{array}\right)}

\begin{document}
\bibliographystyle {unsrt}
\newcommand{\pa}{\partial}
\newcommand{\rhob}{{\bar \rho}}

\titlepage
\begin{flushright}
BGU-PH-95/06, CERN-TH/95-144\\
\end{flushright}
\vspace{19mm}
\begin{center}
{\grande  Relic Gravitational Waves from String Cosmology}

\vspace{10mm}
\centerline{R. Brustein$^{(a,c)}$, M. Gasperini$^{(b)}$, M.
Giovannini$^{(b,c)}$ and G. Veneziano$^{(c)}$}
\bigskip
\centerline{$^{(a)}${\it Department of Physics, Ben-Gurion
University,
Beer-Sheva 84105, Israel}}
\smallskip
\centerline{$^{(b)}${\it Dipartimento di Fisica Teorica,
 Via P.Giuria 1, 10125 Turin, Italy}}
\smallskip
\centerline{$^{(c)}${\it Theory Division, CERN, CH-1211, Geneva 23,
Switzerland} }
\vskip 2  cm

{\medio  Abstract} \\
\end{center}
\noi
A large class of string-cosmology backgrounds leads to
a  spectrum of relic stochastic gravitational
waves, strongly tilted towards  high frequencies,
and characterized by  two basic parameters of the cosmological
model. We estimate the required  sensitivity for  detection
of the predicted gravitational radiation and show that a  region of
our parameter space is within  reach for some of the planned
gravitational-wave
detectors.
\vspace{8mm}

\vfill
\begin{flushleft}
CERN-TH/95-144 \\
June 1995 \end{flushleft}

\newpage
\

\renewcommand{\theequation}{1.\arabic{equation}}
\setcounter{equation}{0}
\section {Introduction}

It is notoriously difficult to find accessible experimental
signatures of
fundamental strings because of their small Planckian size.  Possibly,
an
interesting exception to this rule is represented by the cosmological
predictions of string theory, since these originate from physics of
the
Early-Universe, when space-time curvatures may have been of Planckian
strength.

In order to arrive at some concrete, yet generic,
 predictions of string cosmology, we shall consider a large class
of models in which a period of dilaton-driven inflation
 \cite{Ven,dildriv,GV} is followed by a stringy epoch,
during which the curvature remains of the
order of the string scale $\lambda_s^{-2}$,
and then finally, after possibly a short
dilaton-relaxation era, by the standard (radiation then matter
dominated) cosmology.
As discussed in detail in \cite{BV}, the presence of
a high-curvature stringy epoch appears to be unavoidable for
 a viable inflationary string cosmology scenario. Additional
support to this point of view was given in \cite{KMO}.

Recently, in collaboration with V. Mukhanov \cite{BGGMV},
 we have discussed the main properties of
 metric perturbations in a dilaton-driven background.
Particular attention was given  to the correct treatment of
scalar perturbations
which, in a standard treatment,
appear to grow too large for the applicability of linear
perturbation theory. By carefully ``gauging down" certain growing
modes, we were able to show that both scalar and tensor
perturbations can be treated perturbatively and that they
exhibit very similar spectra. A characteristic feature of these
spectra is that, unlike the spectra of
 the standard inflationary scenarios, they are not
flat but strongly tilted towards  higher frequencies, as originally
noted in \cite{Gasgiov}.

In this note we will  concentrate on tensor perturbations,
 those associated with gravitational waves (GW), at frequencies
which may be accessible to earth-based GW detectors.
Quite amazingly, their spectrum  turns out to be rather independent
of
most of the details of the above string cosmology scenario.
As shown in the first part of this paper,
the spectrum can be completely given in terms of  two parameters, the
value $g_s$ of the string coupling parameter at the end of the
dilaton-driven phase (equivalently, at the beginning of the string
epoch),
and the total red-shift $z_{s}$ occurring during the string era.
The theoretical advantage of considering GW signals stems
 from the fact that, unlike the electromagnetic perturbations
 which underwent a complicated history until recombination,
gravitons decoupled since
 right after the Planck era. As a consequence,
their present spectrum should
be a faithful portrait of the very early Universe.
On the other hand, the detection of  a cosmological background of GW
 requires extremely precise
length measurements, typically at least of the order of
 $\delta L/L<10^{-21}$ \cite{thorne}. In the second part of this
paper we
will show that the expected range of $g_s$ and $z_s$
includes a region which should be accessible
to  future gravitational wave experiments.

Throughout this paper we shall be working in the so-called
String-frame, in which weakly interacting strings
move along geodesic surfaces. Identical results would follow
by adopting the more conventional Einstein frame in which the
curvature is
canonically normalized, but we believe the physics to
be more transparent in the former frame. In the String frame the
string length
parameter which is  the short-distance  cut-off
of string theory,
$\lambda_s = \sqrt{\alpha' \hbar}$, is constant, while the
 Planck length, $\lambda_P = \sqrt{G_N \hbar}$, evolves in time as
$\lambda_P =  e^{ \phi/ 2}\lambda_s $
in a time-dependent dilaton background. In our scenario,
the background evolution
starts from the string perturbative vacuum \cite{GV},
therefore $\lambda_P$ grows from a very small  initial value  to a
value
$g_s\lambda_s $ reached at the beginning of the stringy
era and, finally, to its present (very large!) value of about
$10^{-33}$ cm
at the beginning of the radiation dominated era.

\renewcommand{\theequation}{2.\arabic{equation}}
\setcounter{equation}{0}
\section{Primordial gravitational-wave spectra}

Let us consider, for the moment,   an isotropic,
$(3+1)$-dimensional, spatially flat cosmology. Following
\cite{MGMG} (see also \cite{BGGMV}) it is easy to show
that the canonical variable $\psi$ associated with tensor
perturbations is related to the String-frame metric
$g_{\mu\nu}$ via \bq
g_{\mu\nu} = a^2 (\eta_{\mu\nu} + h_{\mu\nu}) =
a^2 (\eta_{\mu\nu} + {g\over a}\psi_{\mu\nu}),
\label{can}
\eq
where $a(t)$ is the isotropic scale factor,
 $\eta_{\mu\nu}$ is the flat Minkowski metric and  $g =
\exp(\phi/2)$.
The Fourier modes of  each
of the two physical, transverse-traceless, polarizations satisfy
the following simple wave equation \cite{MGMG}
\bq
\psi_k '' + [k^2 - V(\eta)] \psi_k =0,~~~~~~~~~~~~~~~~~V(\eta) =
(g/a)(a/g)''
\label{pert}
\eq
where a prime denotes differentiation with respect to conformal time
$\eta$ ($a d\eta \equiv dt$) and $k$ is
the comoving wave number, related to the physical one, $\omega$, by
$ k = \omega  a$.
Note that, since $a/g$ has power-like behavior in $\eta$ during
the dilaton-driven phase  \cite{Ven,dildriv,GV},
 $V(\eta)$ grows like $\eta^{-2}$
during that epoch reaching a maximum at $\eta = \eta_s$,  i.e.
when $H^{-1} \sim \eta_s a_s \sim \lambda_s \equiv
M_s^{-1}$. We expect $V(\eta)$ to keep growing during the
stringy era and then to fall rapidly to zero at the onset
($\eta=\eta_1$) of the
radiation-dominated  era since, in that phase,
 $a/g \sim \eta$.

 A given mode $k$  will be  well inside the horizon initially,
then hit the potential
barrier $V(\eta)$ at some ``exit" time  $\eta_{ex} \sim k^{-1}$,
and  leave the barrier at  some later ``reentry" time $\eta
= \eta_{re} \sim \eta_1$. The approximate solutions of eq.
(\ref{pert}) in these three regimes are given by
\bq
\psi_k = {\lambda_s \over \sqrt{k}} e^{-ik\eta}\; ,\;
\,\,\,\,\,\,\,\eta < \eta_{ex}
\label{sol1}
\eq
\bq
\psi_k = {a\over g}\left[A_k + B_k \int ^{\eta} d \eta' \left(g
\over a\right)^2\right] \;
,\;\,
 \eta_{ex} < \eta < \eta_{re}
\label{sol2}
\eq
\bq
\psi_k = {\lambda_s \over \sqrt{k}}\left[ c_+ (k) e^{-ik\eta}
+ c_- (k) e^{ik\eta}\right] \; ,\; \,\,\,\eta > \eta_{re}
\label{sol3}
\eq
Equation (\ref{sol1}) enforces the proper normalization
of the primordial vacuum fluctuations.
In the  regime described by eq. (\ref{sol2}) the
perturbation is frozen outside the horizon and the two terms
 appearing there correspond  to the freezing of  $h$ and of its
canonically conjugate momentum, respectively.
Finally, in (\ref{sol3}), the
magnitude of $c_-$ (so-called Bogoliubov coefficient) gives the
amplification of the GW with respect to  a minimal vacuum
fluctuation. The actual value of $|c_-|$ can be easily obtained
 by matching the above solutions and their first derivatives
 at each transition time,
\bq
2 k \eta_1 |c_- (k)| \simeq {g_{ex}/a_{ex} \over g_{re}/a_{re}}
\left[1+ k \eta_1 \left(g_{re}/a_{re} \over
g_{ex}/a_{ex}\right)^2 + k (g_{ex}/a_{ex})^{-2}
\int_{\eta_{ex}}^{\eta_1} d\eta\;(g/a)^2\right] \label{Bog}
\eq

At this point we have to insert some information about the
background evolution. In the simple case at hand of a $D=3+1$
isotropic
 cosmology with static extra dimensions, the
dilaton-driven inflationary background is simply given by
\cite{GV}
\bq
a(\eta) = (-\eta)^{-{1 \over 1+ \sqrt3}} , \;\;\,\,\,\,\,
\phi(\eta) = - \sqrt3 \ln (-\eta) \; , \;\;\,\,\,\,\, a/g \sim
(-\eta)^{1/2}, \,\,\,\,\, -\infty <\eta <0 \label{back}
\eq
while, for the string era, we will assume that $H$
and $\partial_t{\phi}$ are approximately  constant and of order
$\lambda_s^{-1}$. Limiting our attention for the time being
to those scales which crossed the horizon  during the dilaton-driven
phase,
we thus arrive at the following  estimate
 \bq
2 k \eta_1 |c_- (k)| \simeq(k/k_s)^{1/2} z_s (g_s/
 g_1) \left [1+ z_s^{-3}
(g_1/g_s)^{2}+  \ln (k_s/k) + I\right], \,\,\,\,\,\,\, k<k_s
\label{Bog1}
\eq
where we have denoted for convenience
$a_{re}/a_{s}\simeq a_1/a_s=z_s$. In eq.(\ref{Bog1})
 $k_s^{-1}\sim \eta_s\sim (H_sa_s)^{-1}$ is the last
scale exiting during the dilaton-driven phase,
$g_1=\exp(\phi_1/2)$ is a number of order  unity which may be
determined in
term of the (known) present value of the ratio
$\la_p/\la_s$ \cite{K}, and $I$ is the $k$-independent
quantity
\bq
 I = \int_1^{{\eta_1 / \eta_s}} {d\eta\over\eta_s}
{(g/a)^2\over(g_s/a_s)^2} \sim 1+ z_s^{-3}
(g_1/g_s)^{2}\label{int}
\eq
The r.m.s.  perturbation amplitude over a comoving length
scale  $k^{-1}$ (see for instance \cite{BGGMV})
is given, in general, by
$\left|\d {h_k}(\n)\right| \simeq k^{3/2} |h_k| = (g / a)
k^{3/2} |\psi_k|$.
For $\eta>\eta_1$, we then find
\bq
 \left|\d {h_k}(\n)\right| \simeq
{H_1 a_1\over a(\n) M_p} \left(k \over k_s \right)^{1/2}
{g_s\over g_1} z_s
 \left[ 1+{1\over 2}\ln \left(k_s \over k \right)
 + z_s^{-3} \left(g_1\over g_s\right)^{2}\right]
\label{deltah}
 \eq
Equation (\ref{deltah}) is the main result of this section and we
will use it
subsequently to estimate the required sensitivity for detection.

It is worth stressing that the leading
term in $|c_-|$ comes from the integral appearing in
eq. (\ref{sol2}), associated with the freezing of the momentum
variable
conjugate to $h$. This unusual result is due to the fact \cite{GV}
 that, in the Einstein frame, the scale factor is $a_E=a/g$, and
 our background corresponds to a contracting,  rather
than  expanding, Universe. The amplification of
tensor perturbations in a contracting Universe was first
considered long ago in \cite{gris,star}.

In spite of the presence of a high-curvature regime,
we expect our estimates to be valid for scales that went out of the
horizon in the dilaton-driven phase, since they follow from
 the general
physical principle that a perturbation and its  canonically conjugate
momentum should remain frozen while outside the horizon. The
perturbations  thus evolve in a purely kinematical way, giving
rise to the logarithmic term in (\ref{Bog1})  from  evolution  during
 the dilaton-dominated phase and to the second term $I$
  from the stringy epoch.

Finally, let us digress a moment to show
the stability of the result (\ref{deltah}) with respect to
$O(d,d)$ transformations  which connect \cite{MV} different
homogeneous string cosmologies. A simple derivation of this nice
feature
is obtained by working in cosmic  time and by using
directly the amplitude $h$ [defined in eq. (\ref{can})] rather
than the
canonically normalized perturbation $\psi$.
It is straightforward to check that, in  arbitrary $O(d,d)$-related
Bianchi I-type backgrounds (including possible dynamical internal
dimensions and an antisymmetric
tensor $B_{\mu\nu}$), the Fourier modes of $h$
satisfy the following simple equation (see also \cite{MGMG})
\bq
\ddot{h_{\omega}}  - \dot{\bar{\phi}}
 \dot{h_{\omega}} + \omega^2 h_{\omega} = 0
\eq
where dots denote derivatives with respect to
cosmic time and $\bar{\phi} = \phi - \ln |det
g_{\mu\nu}|^{1/2}$
 is the so-called shifted
dilaton, invariant under $O(d,d)$ transformations.
Using the fact that, in any dilaton-driven vacuum
cosmology, $\bar{\phi} \sim
- \ln t$, we easily obtain, for perturbations well outside
the horizon,
\bq
 {h_{\omega}}(t)\sim \ln \left|t\over t_{ex}\right|  \sim \ln
|\omega t| \; ,\;
\eq
showing that the spectrum of GW is  independent
of the chosen string-cosmology background, and increasing our
confidence
that eq. (\ref{deltah}) is indeed  the generic GW spectrum
of a large class of string cosmology models.  Note, incidentally,
that this is not the case for the electromagnetic
perturbations discussed in
\cite{photons}.

\renewcommand{\theequation}{3.\arabic{equation}}
\setcounter{equation}{0}
\section{Observability}

In order to discuss the observability of our signal
it is useful to  rewrite our main result (\ref{deltah})
in terms of present, red-shifted proper
frequencies $\om=k/a$. One easily finds
\bq
\left|\d {h_{\omega}}\right|  \simeq
\sqrt{H_0\over M_s} z_{eq}^{-1/4} g_s z_s
 \left(\omega \over \omega_s\right)^{1/2}
 \left[ 1+{1\over 2}\ln \left(\om_s \over \om \right)
 + z_s^{-3} \left(g_1\over g_s\right)^{2}\right], \,\,\,
\om<\om_s
\label{deltaho}
\eq
\bq
\omega_s = k_s/a \simeq z_{eq}^{-1/4}\sqrt{H_0 M_s}
z_s^{-1} \equiv  z_s^{-1} \omega_1  \sim z_s^{-1} g_1^{1/2}
10^{11} Hz
\label{omegas}
\eq
where  $z_{eq} =a/a_{eq}\sim 10^{4}$ takes into account the
transition from radiation to matter dominance at $t=t_{eq}$,
$\omega_1 =H_1a_1/a \sim 10^{11}$Hz is the maximal
frequency reached during the string phase, $M_s=\la_s^{-1}
\sim H_1$, and $H_0\sim 10^{-18}$Hz is the present value of
the Hubble scale.

It is also convenient to rewrite our results in terms of
another commonly used quantity, the fraction
of critical density, $\Omega_{GW}=\rho_{GW}/\rho_c$, stored in
our GW per logarithmic interval of $\omega$. Defining
$d\Om_{GW}/(d\ln \om)=\om^4|c_-|^2/(M_pH)^2$ we have
\bq
{d \Omega_{GW} \over  d \ln \omega} =
z_{eq}^{-1} g_s^2
 \left(\omega \over \omega_s\right)^3
\left[ 1+{1\over 2}\ln \left(\om_s \over \om \right)
 + z_s^{-3} \left(g_1\over g_s\right)^{2}\right]^2
\sim \left(\om \over H_0\right)^2 \left|\d
{h_{\omega}}\right|^2, \,\,\, \om<\om_s
 \label{Omega}
\eq
It emerges from eq. (\ref{Omega}) that $\omega_s$ plays
 the role of an effective temperature in the sense that,
 below $\omega_s$,
the spectrum is Planckian (up to logarithms of $\omega$).
 The normalization
of the spectrum, however, is different from Planck's because of the
further amplification due to the stringy phase. Also,
 we do not expect the spectrum to stay
Planckian above $\omega_s$ (see below),
but rather to keep growing
 and to reach its maximum at $\omega_1$ before falling exponentially.

Finally we give, with the appropriate caveats,
the generalization of the above results
 to  frequencies whose exit occurred during the stringy phase,
i.e. to the high frequency part
$\omega_s < \omega < \omega_1$. We are aware of the
possible dangers in using field theoretic methods to discuss
perturbations in this regime. However, in the absence of a full
string theoretic calculation, we shall present our results
as an indication of what a possible outcome might be.
One finds, after straightforward calculations,
\bq
|\d {h_{\omega}}|  \simeq g_1
\sqrt{H_0\over M_s} z_{eq}^{-1/4}
\left [\left(\omega \over \omega_1\right)^{2-\beta} +
\left(\omega \over \omega_1\right)^{\beta -1}\right]
\label{delt}
\eq
\bq
{d \Omega_{GW} \over  d \ln \omega} \simeq g_1^2
z_{eq}^{-1}
 \left[\left(\omega \over \omega_1\right)^{6 - 2\beta} +
\left(\omega \over \omega_1\right)^{2\beta}\right], \,\,\,\,\,
\om_s<\om < \om_1
 \label{Omegas}
\eq
where $\beta = - \log( g_s/g_1) / \log z_s$ is also the average
value of $\dot{g}/(g H)$, which we have assumed to vary little
during the string phase. We have also used the fact that during
the string phase the curvature stays controlled by the string
scale $\la_s$ so that, in the String frame, the metric describes
a de Sitter-like expansion with $z_s=a_1/a_s=\eta_s/\eta_1$
(see \cite{Gas} for further details, and for a different
derivation of the same spectrum in the Einstein frame).

We would like  to discuss now the prospects of observing our spectrum
in
gravitational wave detectors. Our main emphasis will be on the
planned
large interferometers LIGO \cite{LIGO} and VIRGO \cite{VIRGO},
 which are expected \cite{thorne} to start operating
at sensitivities (for detection of a stochastic GW background) of
${d \Omega_{GW} /  d \ln \omega} = 10^{-6}$ in a frequency band
around
a few hundred Hz, and have set the ambitious final sensitivity  goals
of
  ${d \Omega_{GW} /  d \ln \omega} = 10^{-10}$
in a frequency band around  $ \omega_I =100 Hz$.  It may well be,
especially in the
first stages of operation, that  coincidence experiments between
bars and interferometers \cite{As}
could also be able to reach similar sensitivities
at frequencies around 1 KHz.
We will mention later other
possible devices which seem to have some good
potential sensitivity,
especially in the higher frequency range, but which have not yet
matured into  concrete operating systems. Detection of
 stochastic GW backgrounds
at frequencies below 1 Hz does not seem  accessible with
current technologies,
and we therefore limit our attention to the range above 1 Hz.

We are interested in finding the regions in our $\{z_s, g_s\}$
parameter space that may be accessible
to experimental detection. From eq. (\ref{omegas}) we can
immediately see that
the accessible region requires large values of $z_s$.
We may distinguish values of $z_s$ in the range $z_s<10^9$
(i. e. $\om_s>\om_I$), in which
the observable spectrum at $\om_I$ comes mainly from perturbations
that
crossed the horizon during the dilaton-driven era,
from those in the range $z_s>10^9$ ($\om_s<\om_I$) in
which the observable spectrum comes mainly from those perturbations
that crossed the horizon
during the stringy era. The predictions in the range $z_s>10^9$
should be considered as less robust than those in the range
$z_s<10^9$.
In addition, we may distinguish
values of $g_s$ in the range $g_s\laq g_1$ in which the dilaton does
not
change much during the stringy phase, from those in the range
$g_s\ll g_1$ in which the dilaton changes by
a large amount during the stringy epoch.

In all cases we have to impose  the bound following
from pulsar-timing measurements \cite{Stin}, which implies
$
{d \Omega_{GW} /  d \ln \omega} \laq 10^{-6}
$
at $\omega_P = 10^{-8} Hz$. We also accept the
 bound $\Om_{GW}\laq 0.1$, imposed by standard nucleosynthesis
\cite{ns}.
 Moreover, we have derived the
spectrum  in the linear approximation, expanding
around a homogeneous background.
We have thus to impose, for
consistency, that the amplified perturbations have a
negligible back-reaction
on the metric, namely $d\Om_{GW} /d\ln  \om <1$ at
all frequencies and times. This, together with the nucleosynthesis
bound, requires a range of parameters corresponding to a
spectrum which is growing also in the stringy phase, $0<\beta<3$,
 namely $(g_s/g_1)<1$ and $(g_s/g_1)>z_s^{-3}$.

Inserting the appropriate numbers in eqs. (\ref{Omega}),
(\ref{Omegas})
 we find, respectively, the following conditions
for detectability in interferometers, i. e.
$
{d \Omega_{GW} /  d \ln \omega} > 10^{-10}
$
at $\omega_I = 10^{2} Hz$
(assuming that the design goals would actually
be achieved),
\bq
 z_{s}^3 g_s^2 \left[11 -{1\over 2} \ln z_s
 + z_s^{-3} (g_s/g_1)^{-2}\right]^2 > 10^{21},
\label{bound1}
\eq
for $z_s < 10^9$,  and either
\bq
\log_{10} {g_1 \over g_s} < \left({1\over 3} +{1\over 18}
\log_{10}g_1^2 \right) \log_{10}z_s, ~~~~~~~~~ \b <3/2
\label{bound2}
\eq
or
\bq
\log_{10} {g_1 \over g_s} > \left({8\over 3} -{1\over 18}
\log_{10}g_1^2\right) \log_{10}z_s, ~~~~~~~~~ \b >3/2
\label{bound3}
\eq
for $z_s > 10^9$.

It may be useful to list approximate forms of the
GW spectral distribution, $d \Om_{GW}/d\ln \om$, in different regions
of parameter space, which we do in {\bf Table 1},
\begin{center}
\begin{tabular}{|r||c|c|} \hline
  &$z_s<10^9$
  & $z_s>10^9$  \\ \hline\hline
 $\b < 3/2$ &$ {z_{eq}^{-1} g_s^2\left(\omega \over \omega_s\right)^3
} $
        &${z_{eq}^{-1} g_1^2\left(\omega
\over \omega_1\right)^{2\beta} }  $
        \\ \hline 
 $\b >3/2$ &${z_{eq}^{-1} z_s^{-6} g_1^4
g_s^{-2}\left(\omega \over \omega_s\right)^3 }$
        &${z_{eq}^{-1} g_1^2
\left(\omega \over \omega_1\right)^{6-2\beta} } $ \\ \hline
\end{tabular}  \end{center}

\noindent
{\small {\bf Table 1.}
 Approximate forms of $(d\Om_{GW}/d\ln \om)$
in various regions of parameter space, $ \w_s= \w_1/z_s$,
$\beta = - \log( g_s/g_1) / \log z_s$.}

\noi
Actually, if one considers also the amplification of
electromagnetic (EM) perturbations in this
scenario \cite{photons}, one finds  an even stronger bound following
from the condition $\Omega_{EM} <1$,
i.e. $(g_1 / g_s) < z_s^2$. Such a constraint
washes out completely the allowed region of parameter space
corresponding to the lower row of {\bf Table 1}. The condition
$\Omega < 1$ has to be imposed, however, for the validity of the
linear
approach, but it could be evaded in a more general, inhomogeneous
model
of background, in which the evolution of fluctuations is treated
non-perturbatively.

In {\bf Figure 1} the allowed region
of parameter space, corresponding to the possible detection of the
spectrum appearing in
the upper row
of {\bf Table 1}, and compatible with the various bounds on
the parameters, is plotted
by taking $g_1=1$ as a reference value. Also shown are the
  parameter intervals in which other detectors may
be useful, for the range
 corresponding to the upper left corner of {\bf Table 1}.

\vspace{.2in}

\centerline{\epsfxsize=5in\epsfbox{gwfig1.epsf}}

\noindent
{\small {\bf Figure 1.}
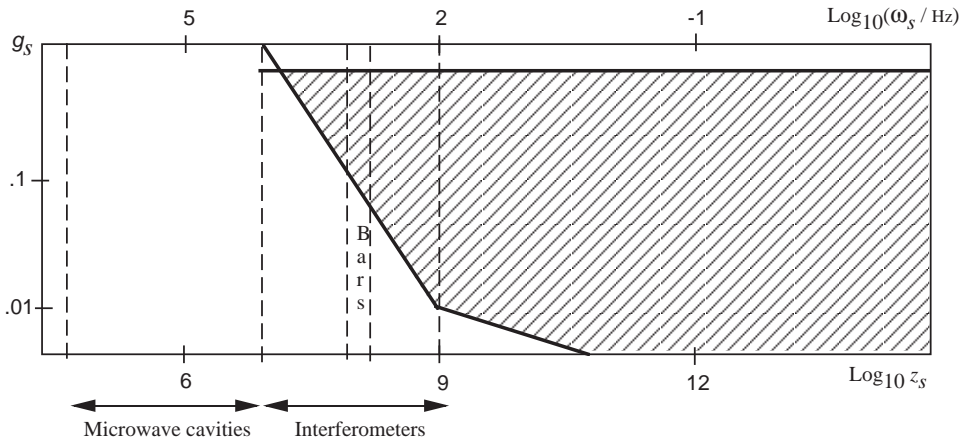
The allowed region in $\{z_s,g_s\}$ parameter space corresponding to
the
first row in {\bf Table 1} is the shaded region
defined by the various constraints.
The dashed lines mark the regions in
which various detectors may be useful,
for the range $z_s<10^9$ and $\b <3/2$.}

We now turn to  discuss in more detail the
observability of our
spectrum as function of frequency, limiting our attention,
for sake of simplicity, to the frequencies leaving the horizon during
the dilaton-driven phase  [eqs.(\ref{deltaho}), (\ref{Omega})].
This spectrum is  pictorially  described  in {\bf Fig. 2} for the
case $\b <3/2$, in terms of the
quantity $|\delta {h_{\omega}}|$ (denoted $h_c$ in
\cite{thorne}), which represents the characteristic amplitude
of a stochastic background.
The odd-shaped region in {\bf Fig. 2} shows detection sensitivities
for the so called ``Advanced LIGO" project,
in terms of the quantity $h_{3/yr}$ defined as the amplitude
necessary for detection of a stochastic background at the 90\%
confidence level
in a 1/3 of a year (see \cite{thorne} for exact definitions).
In {\bf Fig. 2} we can observe clearly  that larger amplification
goes together with larger red-shift for this region of parameter
space.
For a given red-shift $z_{s}$, the higher amplitudes, for all regions
of parameter space, are at higher frequencies.

In addition to interferometers and bars, microwave cavities
may be operated as gravity wave detectors
for the high frequency range $10^6-10^9$ Hz. For the MHz
range specific suggestions \cite{picasso,caves} were actually
implemented \cite{reece}. As can be seen from {\bf Fig. 2}, the
required sensitivity for detection of gravity waves in the MHz
region is $|\delta h_\w|
\sim 10^{-26}$, corresponding to $h_{3/yr}$ of the
same order. We leave to experts to study whether
or not such a sensitivity is accessible with current technologies
and may be reached in a near future, with
microwave cavity detectors or with other experimental
apparatus.

\vspace{.2in}

\centerline{\epsfysize=3.5in\epsfbox{gwfig2.epsf}}

\noindent
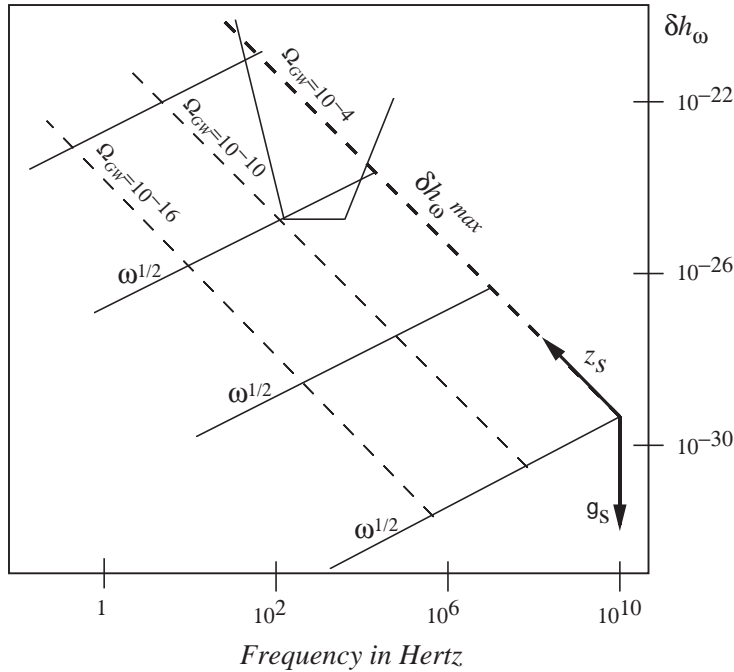
{\small {\bf Figure 2.} The characteristic spectral amplitude
 of gravitational waves $|\delta {h_{\omega}}|$. The solid lines
 show several individual spectra for different values of $z_{s}$
 and $g_s=1$. The thick dashed line shows the maximum
amplitude $|\delta h_\omega^{max}|$ as a function of $z_{s}$
for $g_s=1$. The dashed lines are lines of fixed $g_s$
and therefore lines of constant energy density. $\Omega_{GW}$ is
roughly the maximal amount of gravitational energy density at a given
value of $g_s$. Also shown in the figure is the odd-shaped region
marking the sensitivity goals for the detection of
a stochastic background according to
 ``Advanced LIGO" project.}

\renewcommand{\theequation}{4.\arabic{equation}}
\setcounter{equation}{0}
\section{Conclusion}

We showed that a rather generic string cosmology
scenario leads, naturally, to the production of an amplified
quasi-thermal spectrum of gravitons during the dilaton-driven
phase. This spectrum is very stable under modifications of the
background and in particular under $O(d,d)$ transformations.
The slope of the spectrum may change for modes crossing the
horizon in the subsequent string phase, but remains in general
characterized by an enhanced production of high frequency
gravitons, irrespective of the particular
value of the spectral index.

We showed, in particular, that it may be possible to
detect  such a relic GW background with large interferometers
for a range of the two parameters characterizing our class of
models. We would like, however,
to encourage the study and the developments of gravitational
detectors with enhanced sensitivity in the high frequency,
KHz - GHz, range.   This frequency band should be in fact all but
a ``desert" of relic gravitational radiation that one may
expect on the grounds of the standard inflationary
scenario or from ordinary
astrophysical sources. Our string cosmology scenario
is unique in predicting a
strong signal in this range of frequencies. In general, a
sensitivity of $\Om \sim 10^{-4} - 10^{-5}$ (which is not out of
reach,
in the KHz region, for coincidence experiments between bars and
interferometers \cite{As}),
could be already enough to detect a signal, so even a null
result at that level of sensitivity would already constrain in a
significant way the parameters of the string background, while
detection
would provide a first glimpse at some new and exciting Planckian
physics.

As we stressed, many of our results are independent of details of the
string cosmology scenario. However, it would be worthwhile
pointing out again that, although some ideas have been put forward
\cite{ideas},
a solid string-theoretic treatment of the
stringy phase, which we propose as  the necessary bridge
between the dilaton-driven and the standard decelerated era,
does not yet exist.
Recent progress \cite{ideas}
on the interpretation of singularities in
string theory, as simply a failure to describe physics in terms of
the original set of massless fields, may shed light on the
resolution of this issue. The
understanding of singularities in string theory
would certainly help putting our string cosmology scenario on a
firmer
basis, and may even provide a framework for the calculation of
the parameters $g_s$ and $z_s$.

\vskip 1.5 cm
\noi
{\bf Acknowledgements}\\

\noi
R. B. is supported in part by an Alon Grant. We would like to thank
E. Coccia,
S. Finn, P. Michelson,  P. Saulson and  N. Robertson for discussions
about gravity wave detectors.

\end{document}